# AMC Model for Denial of Sleep Attack Detection

**[1]Tapalina Bhattasali, [2]Rituparna Chaki**
[1]Techno India College of Technology, Kolkata, India
tapolinab@gmail.com
[2]West Bengal University of Technology, Kolkata, India
rituchaki@gmail.com

**ABSTRACT.** Due to deployment in hostile environment, wireless sensor network is vulnerable to various attacks. Exhausted sensor nodes in sensor network become a challenging issue because it disrupts the normal connectivity of the network. Affected nodes give rise to denial of service that resists to get the objective of sensor network in real life. A mathematical model based on Absorbing Markov Chain (AMC) is proposed for Denial of Sleep attack detection in sensor network. In this mechanism, whether sensor network is affected by denial of sleep attack or not can be decided by considering expected death time of sensor network under normal scenario.

## INTRODUCTION

In wireless sensor network, power consumption of the wireless radio is the dominating factor of the battery life of the sensor nodes. The nodes normally enter a low power sleep state to conserve energy and extend their battery life. Most devastating Denial of sleep attack [1] prevents the sensor nodes from entering into low power sleep state, thereby draining their batteries at much higher rates than normal usage. The nature of deployment makes the network almost impossible to recharge or replace the battery power of sensor nodes. Quick battery depletion leads to sudden death of sensor nodes which is responsible for death network either fully or partially. As a consequence transmission of critical data is compromised. This may give rise to emergency situation.

In denial of sleep attack, goal of the intruder is to maximize the power consumption of the sensor node, thereby decreasing its battery life. The attack achieves this by keeping the sensor node busy and preventing it from going into low power sleep mode. An attacker can exhaust a node's resources by repeatedly sending RTS messages triggering CTS responses from a targeted node. In this case, all the nodes within the radio range of the sender will be receiving those (RTS) control packets, thus draining their power supplies. The attacker may also send a constant stream of unauthenticated or replayed broadcast packets causing the nodes to remain awake.

Most of the existing detection mechanisms are based on deterministic model. But in sensor network, status of the affected nodes may be changed with time. The need of the day is to propose a probabilistic model for accurately detect attack.

Absorbing Markov Chain (AMC) is useful [2] to detect the probabilistic nature of sensor nodes. Sensor nodes behavior or state can be changed with respect to time. In an absorbing Markov chain, the probability that the process will be absorbed when no transition is possible from this state to other states is 1. When any process enters into this state it remains in this state forever. This state surely comes at the end of the transition. This nature is almost similar to the dead state of sensor network when batteries of sensor nodes are exhausted because of intrusion. Hence the probability of dead state is 1.

According to the mathematical theory of probability, an absorbing Markov chain can be defined as a Markov chain in which every state can reach to an absorbing state. An absorbing state is a state that, once entered, cannot be left. Markov chain is an absorbing chain if there is at least one absorbing state and it is possible to go from any state to at least one absorbing state within a finite number of steps. In an absorbing Markov chain, a state that is not absorbing is called transient.

## PROPOSED METHODOLOGY

To estimate effect of denial of sleep attack, a model is proposed to analyze the behavior of compromised sensor nodes based on the Markov chain with an absorbing state. This model is used to decide whether network is affected by denial of sleep attack or not. Each sensor node behavior can be modeled by Absorbing Markov Chain. Different states of sensor nodes will be represented by the states of Markov Chain.

A sensor node can get one of the states such as sleep state, active state, inactive state, dead state. During the lifetime of sensor nodes following transitions can be possible.

Sleep → Active → Dead.
Sleep → Inactive → Dead.
Sleep → Active → Inactive → Active → Dead.
Sleep → Active → Sleep → Inactive → Active → Dead.

It is assumed that, initial state of sensor node is sleep state from which it can go to either active state or idle or inactive state. From active state it can go to either sleep state or inactive state or dead state. From inactive state sensor node can go to either active state or dead state which is the final state. Following are the definitions of the states of sensor node in its life cycle.

Sleep state can be defined as state S in which a sensor node remains in low power energy saving status.

Inactive state can be defined as state S in which a sensor node remains awake but does nothing. The status of the node is idle or waiting.

Active state can be defined as state S in which a sensor node may sense or transmit data or compute and energy consumption is maximum in this state.

Dead state can be defined as state S in which a battery of sensor node is exhausted and a sensor node is failed.

This proposal mainly focuses on overall network performance. Instead of focusing on single sensor node's behavior, it detects intrusion by monitoring network flow. It ignores the attack till the normal data transmission rate is maintained. It is assumed that network is in dead state when 4/5 of all deployed sensor nodes are in dead state. In this methodology, expected absorption time of sensor network is analyzed which indicates network lifetime. If the network state tends to death fast compared to normal death time of sensor network, it is assumed that network is affected by denial of sleep attack.

### Estimation Of Expected Time Of Death Of Sensor Network

Here the expected time at which the process reaches to absorbing state (dead) needs to be analyzed [3,4]. n step transition probabilities from sleep state to dead state for proposed Markov chain model is defined by,

$Prob_{SLEEP-DEAD}^{(n)} =$
$P | X_n = DEAD | X_0 = SLEEP |$

The dead state is called absorbing if $Prob_{DEAD-DEAD} = 1$. It is assumed that node density of wireless sensor network is N. Once the system reaches to dead state, it stays there forever. Normally, dead state of sensor network can arise when node density i=0 (initially no sensor node) and i=M (threshold value M= (4/5)*N). Sensor network can be considered dead when 80% nodes are in dead state.

In this model, state transition can either from next state or previous state or current state.

When sensor nodes goes from state i to state j = i+1 (current state → next state), transition probability becomes,
$\lambda_i = TProb_{i+1} = ((M - i)/ M) \cdot (i/M)$ .

When process goes from i to j = i-1 (current state → previous state), transition probability becomes,
$\gamma_i = TProb_{i-1} = ((M - i)/ M).(i/M)$.
When system goes from i to j=i (current state → current state), transition probability becomes,
$TProb_i = [((M - i)/ M).((M - i)/ M)] + [(i/M).(i/M)]$.
$TProb_{i,j} = 0$ for all other values of j, since it is impossible to make other transitions.
It is assumed that, $\gamma_i / \lambda_i = \beta_i$ and $\beta_0 = 1$.
Probability of death can be defined as,
$\psi^i = (\sum_{k=0}^{i-1} \beta_k) / (\sum_{k=0}^{M-1} \beta_k)$.
Mean number of times the system is in j state and it is started in state i,
$t_i' = \sum_{j=1}^{M-1} t_{ij}'$
$= \sum_{j=1}^{i} [((1- \psi^i). \sum_{k=0}^{i-1} \beta_k) / (\beta_{j-1} . \gamma_j)] + \sum_{j=i+1}^{M-1} [(\psi^i . \sum_{k=0}^{M-1} \beta_k) / (\beta_j . \lambda_j)]$
In this model, it is seen that,
$\gamma_i = \lambda_i = (i.(M-1))/(M)^2$, therefore $\beta_i = 1$, for i=0,…, M and $\psi^i = i/M$, for M=(4/5)*N.
Expected time of death of sensor network can be determined by as follows.
$T_i' = \sum_{j=1}^{M-1} T_{ij}'$
$= \sum_{j=1}^{i} [((1- \psi^i). \sum_{k=0}^{j-1} \beta_k) / (\beta_{j-1} . \Gamma_j)]$
$+ \sum_{j=i+1}^{M-1} [(\psi^i . \sum_{k=j}^{M-1} \beta_k) / (\beta_j . \lambda_j)]$
Putting values it becomes,
$T_i' = \sum_{j=1}^{i} [(M.(M-i))/(M-j)] + \sum_{j=i+1}^{M-1} (M.i)/j$
$= M.(M-i). \sum_{j=1}^{i} (1/(M-j)) + M.i. \sum_{j=i+1}^{M-1} (1/j)$

This estimation is useful for detecting whether network is affected by denial of sleep attack or not.

If value of M is changed and $T_i'$< normal lifetime of sensor network, then network is affected by denial of sleep attack.

## CONCLUSION

In this paper Absorbing Markov Chain is applied to the proposed model to analyze network lifetime for considering whether sensor network is affected by denial of sleep attack or not. As behavior of sensor nodes varies with time, proposed intrusion detection model in wireless sensor network is probabilistic one. It works more accurately than deterministic model.

In this paper, only proposed methodology is given briefly. Now work is on for analyzing the performance of AMC model for detecting denial of sleep attack by extending the proposed methodology.

## REFERENCES

[1] Stajano F. and Anderson R., "The Resurrecting Duckling: Security Issues for Ad-hoc Wireless Networks," in Proceedings of the 7th International Workshop on Security Protocols, pp. 172–194, April, 1999, Springer-Verlag London, UK, ISBN:3-540-67381-4.

[2] Shakhov V., Popkov V., "Performance Analysis of Sleeping Attacks in Wireless Sensor Networks", International Conference on Computational Technologies in Electrical and Electronics Engineering, IEEE REGION 8 SIBIRCON 2008, Novosibirsk, 21-25 July, 2008, pp. 418-420. ISBN: 978-1-4244-2133-6.

[3] Absorbing states in Markov chains. Mean time to absorption. Wright-Fisher Model. Moran Model, Antonina Mitrofanova, NYU, department of Computer Science , December 18, 2007.

[4] Warren Ewens, "Mathematical Population Genetics", Second edition, 2004, pp 20-23, 86-91, 92-99, 104-109.

# Authors

Tapalina Bhattasali has received her M. Tech. Degree in Information Technology from West Bengal University of Technology. She is at present working at Techno India College of Technology, Kolkata, India. Her research interests include the field of Wireless Sensor Network.

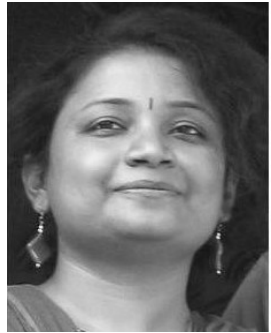

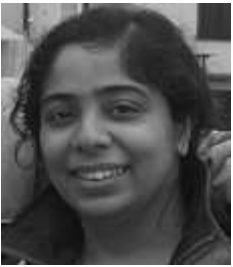

Rituparna Chaki is Associate Professor in the Department of Computer Science & Engineering of West Bengal University of Technology, Kolkata, India since 2005. She received her Ph.D. from Jadavpur University, India. The primary areas of research interest. Dr. Chaki has number of international publications to her credit. Dr. Chaki has also served in the committees of several international conferences.